\documentclass[a4paper,12pt,dvipsnames,usernames]{article}
\pdfoutput=1

\usepackage{aas_macros}

\usepackage{jcappub} 

\usepackage[utf8]{inputenc}
\usepackage{graphicx}
\usepackage{amsmath}
\usepackage{amssymb}
\usepackage{bm}
\usepackage{paralist,array}
\usepackage{units} 
\graphicspath{{fig/}{./Plots/}} 
\usepackage{slashed}
\usepackage{todonotes}
\usepackage[parfill]{parskip}
\usepackage{longtable}
\usepackage{subfig}
\usepackage{float}

\graphicspath{{plots/}}

\newcommand{\lsim}{\raisebox{-0.13cm}{~\shortstack{$<$ \\[-0.07cm]
      $\sim$}}~}

\title{Searching for Synchrotron Emission from the Geminga TeV Halo using the Planck Satellite}

\author[a,b,c]{Dan Hooper,}
\author[a,b]{Elena Pinetti,}
\author[a,b]{and Anastasia Sokolenko}

\affiliation[a]{Fermi National Accelerator Laboratory, Theoretical Astrophysics Department, Batavia, IL, 60510, USA}
\affiliation[b]{University of Chicago, Kavli Institute for Cosmological Physics, Chicago, IL 60637, USA}
\affiliation[c]{University of Chicago, Department of Astronomy \& Astrophysics, Chicago, IL 60637, USA}

\begin{document}

\abstract{Pulsars convert a significant fraction of their total spin-down power into very high-energy electrons, leading to the formation of TeV halos. It is not yet known, however, whether these sources also efficiently accelerate electrons at lower energies and, if so, how those particles propagate through the surrounding environment. If pulsars produce $\sim 50-300 \, {\rm GeV}$ electrons, these particles would produce a spatially extended halo of synchrotron emission in the frequency range measured by Planck. Such emission could be used to constrain the low-energy diffusion coefficient in the regions surrounding these pulsars, as well as the spectrum and intensity of the electrons that are accelerated in this energy range. In this study, we attempt to use Planck data to constrain the nature of the Geminga pulsar's TeV halo. We find no conclusive evidence of this emission in Planck's frequency range, however, and calculate that the synchrotron flux from Geminga should be well below the total flux measured by Planck, even for models with favorable diffusion parameters or soft injection spectra. At this time, these measurements are not capable of significantly constraining the values of these parameters.}

\maketitle

\section{Introduction}

Observations by the HAWC, LHAASO, and HESS Collaborations have shown that young and middle-aged pulsars produce large fluxes of spatially extended emission at multi-TeV energies~\cite{HAWC:2017kbo,Linden:2017vvb,Sudoh:2019lav,Sudoh:2021avj,HAWC:2020hrt,2019ICRC...36..797S,HESS:2018pbp,HESS:2017lee,LHAASO:2023rpg,LHAASO:2021crt,Fang:2021qon,Abeysekara:2017hyn}. The measured intensity of this emission indicates that these pulsars convert a significant fraction of their total spin-down power into very high-energy electron-positron pairs, which produce the observed gamma rays through inverse Compton scattering~\cite{Hooper_2017,Profumo_2018,Fang_2018,DiMauro:2020cbn,Evoli:2020szd, Manconi:2020ipm,Orusa:2021tts}. The resulting TeV halos extend out to $\sim 25 \, {\rm pc}$ in radius, indicating that cosmic rays propagate much less efficiently in the vicinity of these sources than they do elsewhere in the interstellar medium (ISM)~\cite{Hooper_2017,2005yCat.7245....0M,Lopez-Coto:2017pbk,Johannesson:2019jlk,DiMauro:2019hwn,Evoli:2018aza,Fang:2019iym,Hooper:2021kyp,HAWC:2017kbo, Abeysekara:2017hyn, Kappl:2015bqa, Genolini:2019ewc,Profumo_2018, Tang_2019, J_hannesson_2019}.  

There are many open questions regarding the physics of TeV halos. Perhaps most notably, it is not yet clear why diffusion is so slow within the regions that surround these sources~\cite{Schroer:2023aoh,Mukhopadhyay:2021dyh,DeLaTorreLuque:2022chz,Liu:2019zyj,Recchia:2021kty,Bao:2021hey,Evoli:2018aza,Fang:2019iym}. In a previous study, we assessed the ability of the Cherenkov Telescope Array (CTA) to discriminate between various models of TeV halos, focusing on the specific case of the Geminga pulsar~\cite{Hooper:2023mvd}. While observations of TeV halos at and above TeV-scale energies can be quite informative, such measurements do not reveal much about the electrons that these sources generate at lower energies, or how those particles propagate through the surrounding volume of space.

In this paper, we consider the ability of the Planck satellite to detect and measure the synchrotron emission from the GeV-TeV electrons that are expected to be accelerated by TeV halos. For electrons with energies in the range of $\sim 50-300 \, {\rm GeV}$, the resulting synchrotron emission is predicted to peak in the frequency range measured by Planck, providing us with a way of constraining the production and diffusion of electrons in this energy range. We focus in this study on the TeV halo associated with the Geminga pulsar, finding no conclusive evidence of synchrotron emission from this source. This information can be used to constrain the physics behind Geminga's TeV halo.

\section{Radio Synchrotron From TeV Halos}

The spectrum of synchrotron emission from a relativistic electron features a broad peak at approximately one third of the critical frequency~\cite{Blumenthal:1970gc}:
\begin{align}
\nu_{\rm peak} \sim \frac{\nu_c}{3} &= \frac{1}{2} \gamma^2_e \nu_g \sin \alpha_p \\
& \approx 40 \, {\rm GHz} \times \bigg(\frac{E_e}{60 \, {\rm GeV}}\bigg)^2 \bigg(\frac{B}{3 \, \mu {\rm G}}\bigg) \bigg(\frac{\sin \alpha_p}{2/3}\bigg), \nonumber
\end{align}
where $\nu_g =  eB/2\pi m_e$ is the non-relativistic gyrofrequency, and $\alpha_p$ is the pitch angle. From this expression, we expect electrons with energies in the range of $\sim 50-300 \, {\rm GeV}$ to produce synchrotron emission that peaks in the frequency range measured by Planck ($30-857 \,{\rm GHz}$).

In the Thomson regime ($E_e \ll m^2_e/\epsilon_{\gamma}$), the processes of synchrotron and inverse Compton scattering cause electrons to lose energy at the following rate~\cite{Blumenthal:1970gc}:
\begin{align}
\frac{dE_e}{dt} &\approx -\frac{4}{3} \sigma_t \rho_{\rm mag} v_e^2 \gamma^2_e -\frac{4}{3} \sigma_t \rho_{\rm rad} v_e^2 \gamma^2_e  \\
&\approx -4.5
\times 10^{-13} \, {\rm GeV/s} \times \bigg(\frac{E_e}{60 \, {\rm GeV}}\bigg)^2  \,
\Bigg[\bigg(\frac{B}{3 \, \mu{\rm G}}\bigg)^2 + \bigg(\frac{\rho_{\rm rad}}{1 \, {\rm eV/cm}^3}\bigg) \Bigg], \nonumber
\end{align}
where $\sigma_{\rm t}$ is the Thomson cross-section, $\rho_{\rm mag} = B^2/2\mu_0$ is the energy density of the magnetic field, and $\rho_{\rm rad}$ is the energy density of the radiation field. From this expression, we see that electrons with $E_e \lsim 100 \, {\rm GeV}$ lose less than 10\% of their energy over the lifetime of the Geminga pulsar ($t_{\rm Geminga} \approx 340 \, {\rm kyr}$).

Over a time, $t$, an electron is typically displaced via diffusion by a distance on the order of
\begin{align}
L &\sim 2\sqrt{Dt} \approx 40 \, {\rm pc} \times \bigg(\frac{D}{4\times 10^{26} \,{\rm cm}^2/{\rm s}}\bigg)^{1/2} \bigg(\frac{t}{340\,{\rm kyr}}\bigg)^{1/2},
\end{align}
where $D$ is the diffusion coefficient. In this expression, we have neglected energy losses, as justified for the case of electrons with energies below $\sim 100 \, {\rm GeV}$.

The observed angular extent of TeV halos indicates that diffusion is highly suppressed in the regions that surround these sources, in particular when compared to that of the overall ISM~\cite{Hooper_2017, Fang_2018, Profumo_2018, Tang_2019, J_hannesson_2019}. In Ref.~\cite{Hooper:2023mvd}, we found that data from the direction of Geminga, as collected by HAWC~\cite{HAWC:2017kbo} and HESS~\cite{HESS:2023sbf}, could be well fit by a model with $D(E_e) = 10^{26} \, {\rm cm}^2/{\rm s} \times (E_e/{\rm GeV})^{1/3}$, which is a factor of several hundred times smaller than that of the ISM~\cite{Strong:2007nh,Trotta:2010mx}. 

From the above considerations, we expect any $\sim 10 \, {\rm GeV}-{\rm TeV}$ electrons that are accelerated by young and middle aged pulsars to produce a spatially extended flux of synchrotron emission at radio frequencies. Measurements in this frequency range could thus be used to constrain the low-energy diffusion coefficient in the regions surrounding these pulsars, as well as the spectrum and intensity of the electrons that are accelerated in this energy range.

\section{Planck Data and TeV Halos}
\label{sec:planck}

The Planck satellite has measured the sky in 9 frequency bands, ranging from 30 to 857 GHz~\cite{Planck:2018bsf, Planck:2018lkk}. In this study, we use the raw data from the Planck Public Data Release 3 in each of these frequency bands, and compare the flux observed from the direction of the Geminga pulsar (PSR J0633+1746) to that predicted by different models of its TeV halo.

\begin{figure}[t]
    \centering
    \includegraphics[width=0.48\linewidth]{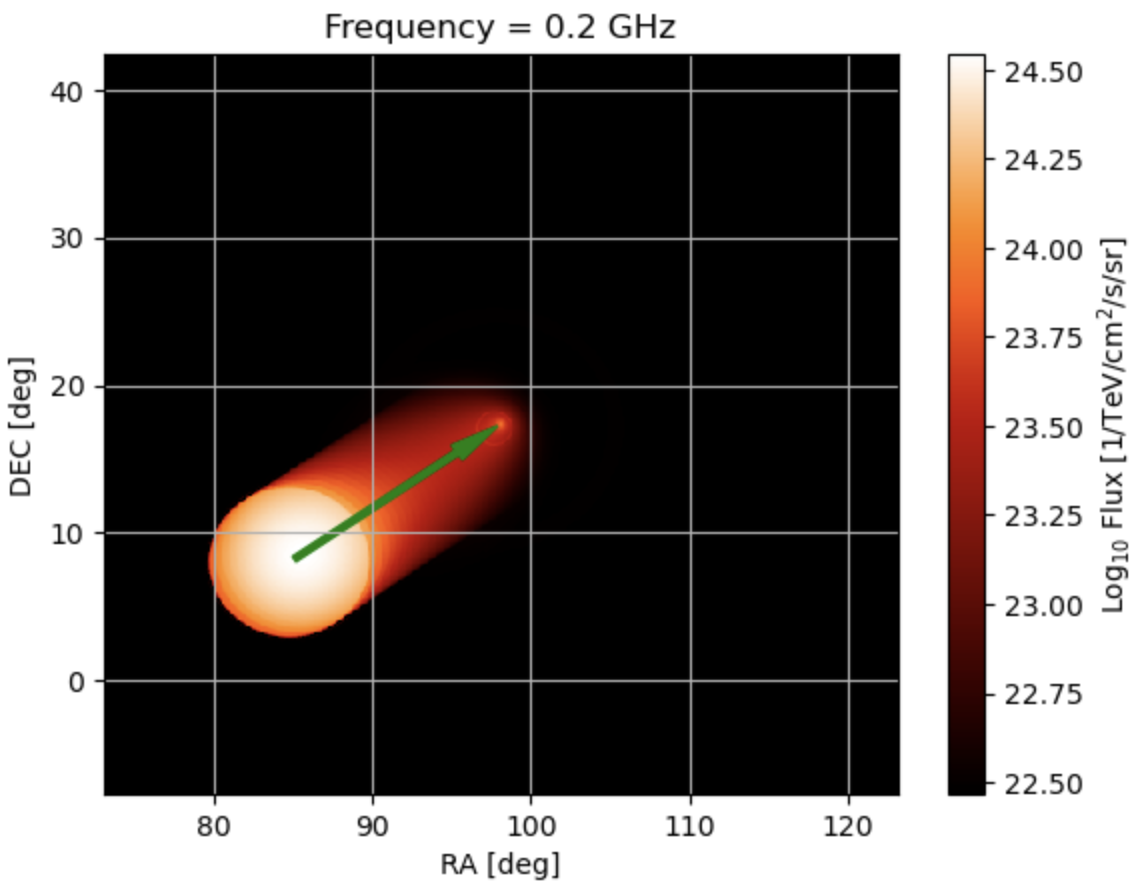}~\includegraphics[width=0.48\linewidth]{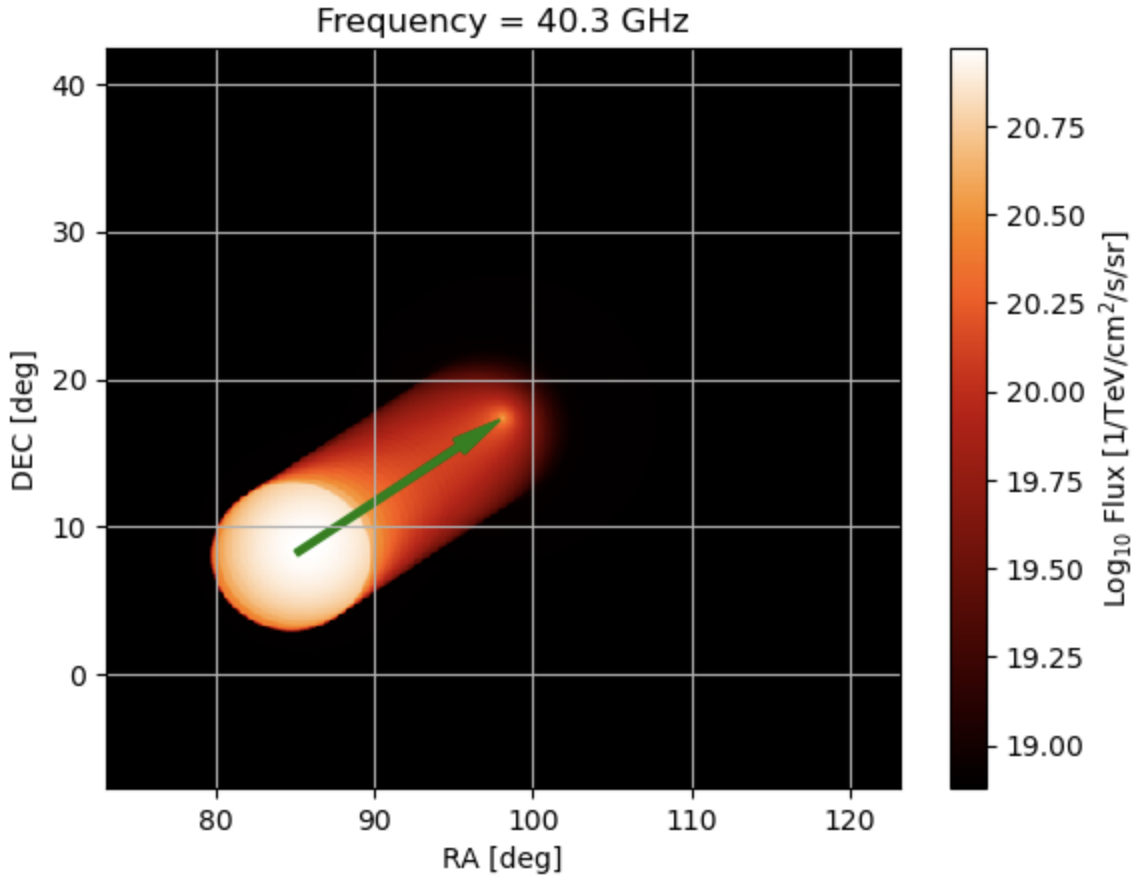} 
    \\
    \includegraphics[width=0.48\linewidth]{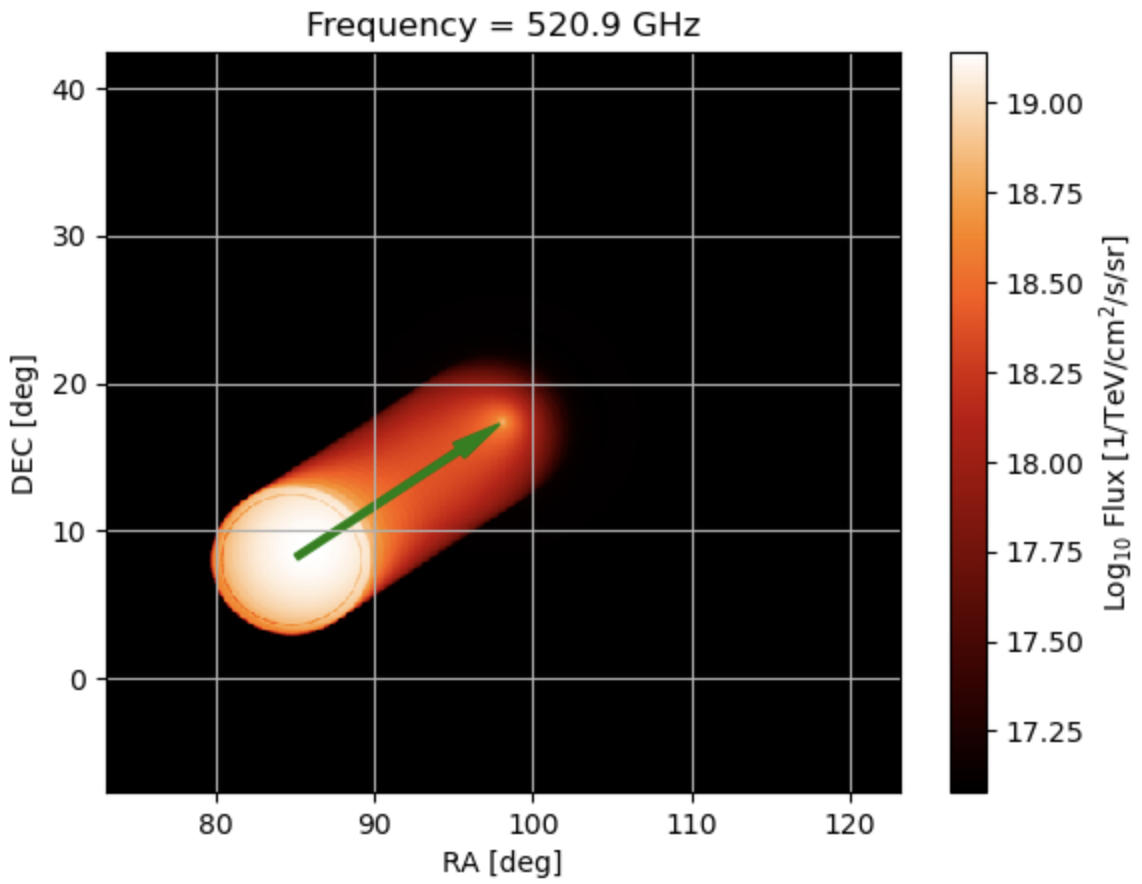}~\includegraphics[width=0.48\linewidth]{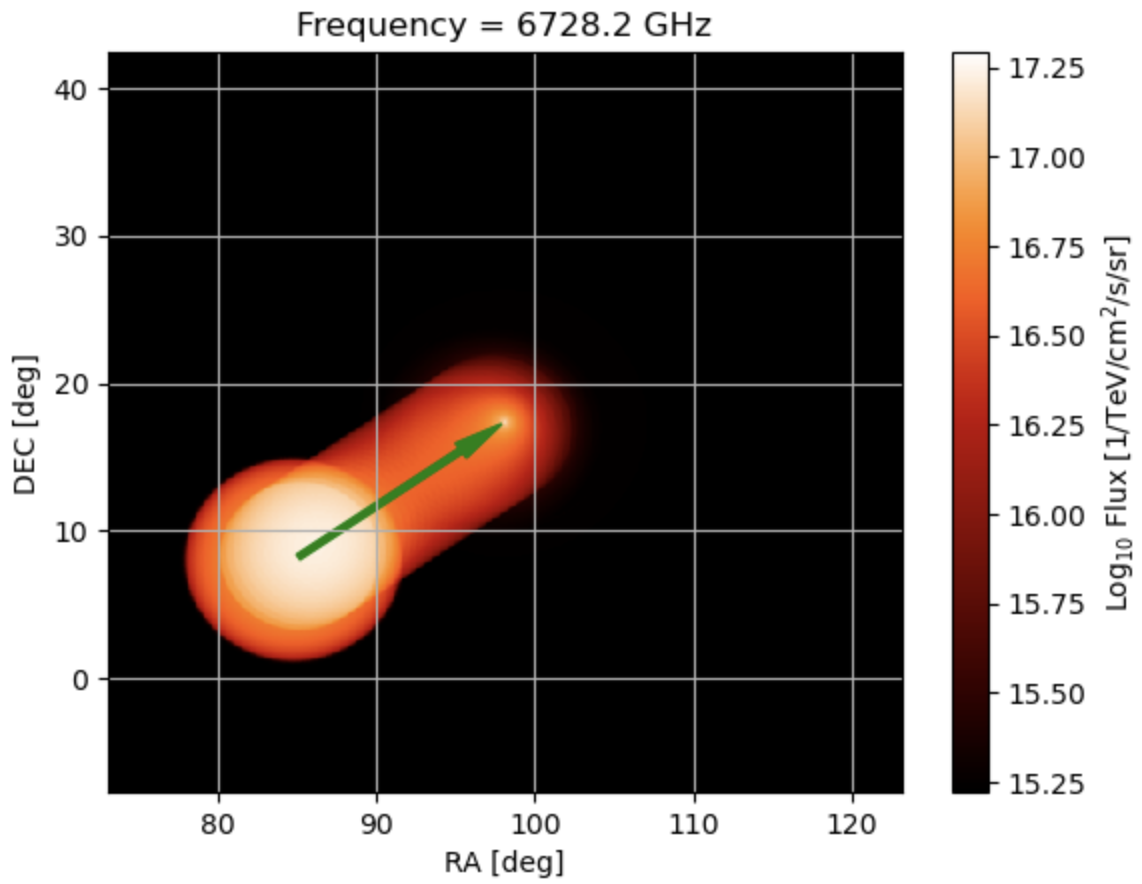} 
    \caption{The intensity and morphology of the synchrotron emission predicted from the Geminga pulsar at four selected frequencies. Here we have adopted our default model for Geminga, as described in Ref.~\cite{Hooper:2023mvd}. The green arrow in each frame indicates the proper motion of Geminga over its lifetime.}
    \label{fig:flux_proper_motion} 
\end{figure}

The Geminga pulsar is measured to have a proper motion of 178.2 mas/yr, which over 340 kyr corresponds to a total displacement of $16.8^{\circ}$~\cite{Hobbs:2005yx}. We assume that the rate of Geminga's proper motion has been constant over its lifetime and that the halo and pulsar have moved together over this period of time. Given the fact that electrons in the energy range of interest do not experience significant energy losses over the lifetime of Geminga, the signal in the Planck frequency range is expected to extend from the location of Geminga's birth to the present location of this pulsar. 

In Fig.~\ref{fig:flux_proper_motion}, we plot the intensity and morphology of the synchrotron emission predicted from the Geminga pulsar at four selected frequencies. In this calculation, we have adopted our default model for Geminga, as described in our previous study~\cite{Hooper:2023mvd}. This model has been shown to provide a good fit to the high-energy emission observed by HAWC~\cite{HAWC:2017kbo} and HESS~\cite{HESS:2023sbf}. The green arrow in each frame indicates the proper motion of Geminga over its lifetime. Note that the brightest flux of synchrotron is not expected from Geminga's present location, but from the angular region surrounding its birth. This is due to the fact that Geminga's spindown power was much greater when it was young than it is in the present era.

\begin{figure}[H]
    \centering
    \includegraphics[width=0.39\linewidth]{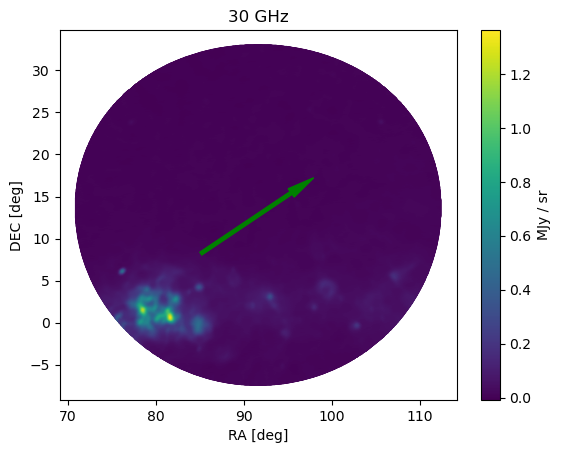}~\includegraphics[width=0.38\linewidth]{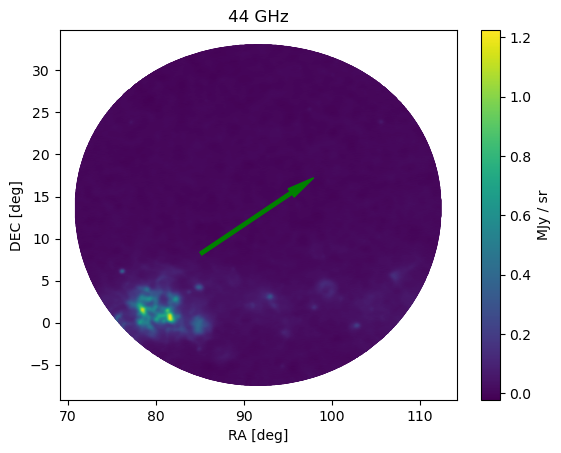}
    \\
    \includegraphics[width=0.38\linewidth]{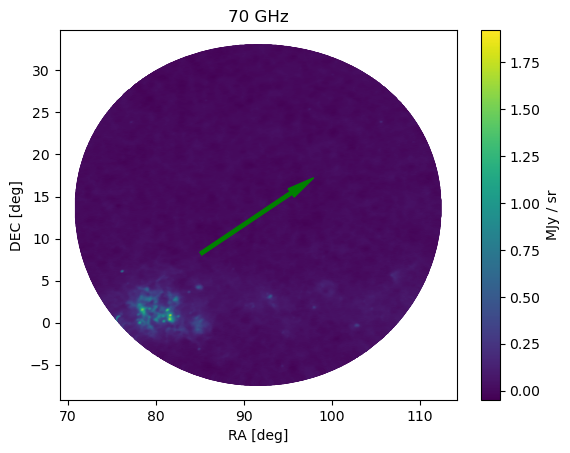}~\includegraphics[width=0.39\linewidth]{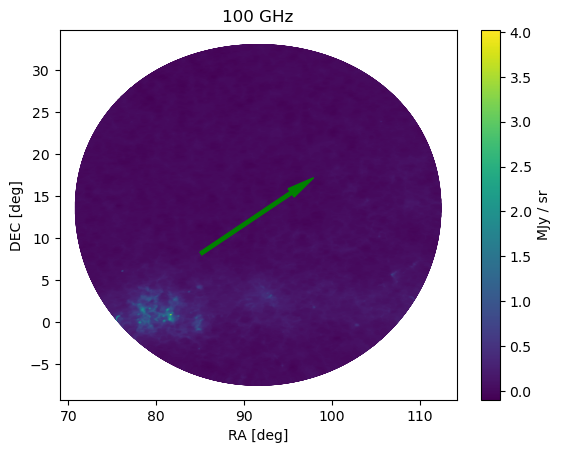}
    \\
    \includegraphics[width=0.39\linewidth]{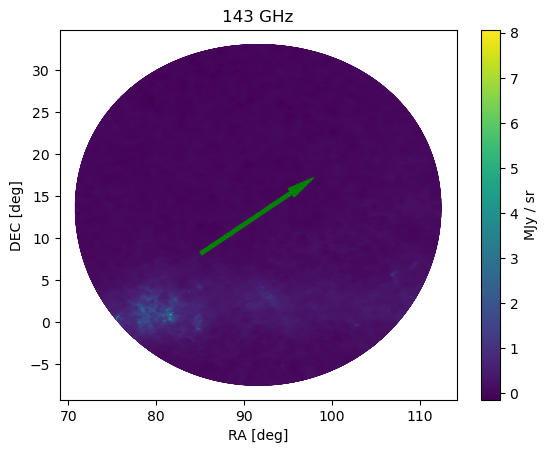}~\includegraphics[width=0.39\linewidth]{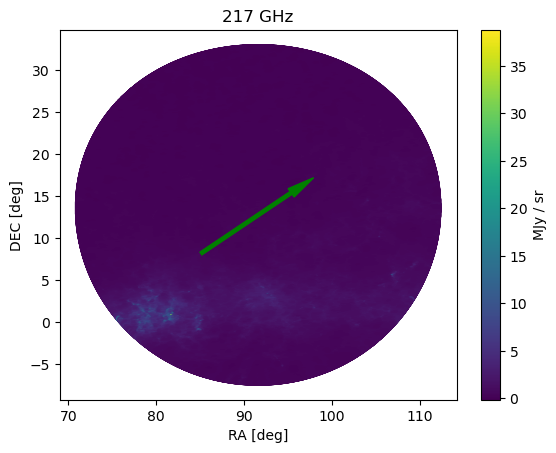}
    \\
    \includegraphics[width=0.39\linewidth]{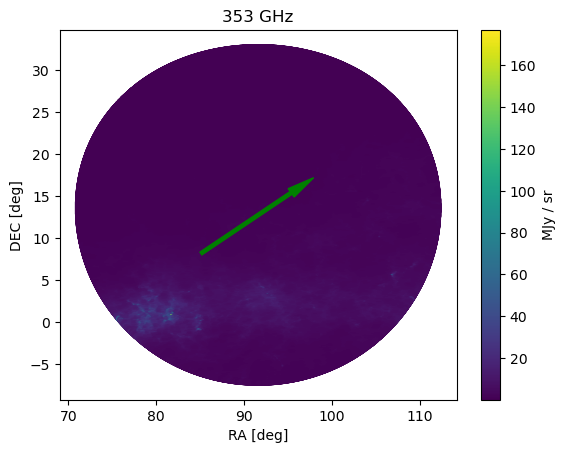}~\includegraphics[width=0.39\linewidth]{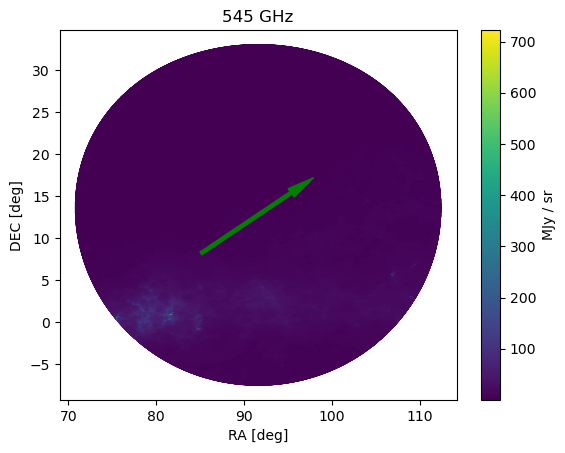}
    \caption{The emission measured by the Planck satellite from the direction of the Geminga pulsar. The arrows indicate the proper motion of Geminga over its lifetime. These frames show no obvious signs of emission from this source. The green arrow shows the proper motion of Geminga.}
    \label{fig:Geminga_20deg} 
\end{figure}

In Fig.~\ref{fig:Geminga_20deg}, we show the raw flux observed by Planck in eight frequency bands in the direction of Geminga. The center of these frames is chosen to be ${\rm RA} = 91.6^\circ$, ${\rm DEC} = 12.8^\circ$, which is the midway point between the birth and current position of Geminga~\cite{HAWC:2017kbo, Abeysekara:2017hyn,Hobbs:2005yx}. These frames show no obvious signs of emission from this source (although these maps do include some emission that could plausibly originate from the region surrounding Geminga's birth, especially at the lowest measured frequencies). In light of this information, we will proceed in the following section to use the lack of synchrotron emission observed from Geminga to constrain our model of this pulsar's TeV halo.

Note that we have not considered the Monogem pulsar (PSR J0659+1414) in this study, despite the fact that it is approximately half as bright as Geminga at multi-TeV energies. The reason for this decision is that Monogem is much younger than Geminga ($t_{\rm Monogem} \approx 110 \, {\rm kyr}$), and thus the total integrated energy of the emission from this source is likely to be much lower than that of Geminga. Under the assumption of magnetic dipole breaking, $\dot{E}(t) \propto [1+(t/\tau)]^{-2}$, and otherwise adopting the same parameters for Geminga and Monogem, we estimate the following ratio for their total time-integrated emission:
\begin{align}
\frac{E_{\rm Geminga}}{E_{\rm Monogem}} \approx \frac{F^{\rm Geminga}_{\rm TeV}}{F^{\rm Monogem}_{\rm TeV}} \times \bigg(\frac{t_{\rm Geminga}}{t_{\rm Monogem}}\bigg)^2 \sim 20,
\end{align}
where $F_{\rm TeV}^{\rm Geminga}$ and $F_{\rm TeV}^{\rm Monogem}$ denote the current fluxes of very high-energy emission from these sources. In light of this comparison, we expect Monogem to be $\sim 20$ times fainter than Geminga in the frequency range measured by Planck.

\section{Results}
\label{sec:results}

In this section, we use a two-zone model for Geminga's TeV halo, with parameters as described in Ref.~\cite{Hooper:2023mvd}, and use the lack of synchrotron emission observed from the direction of this object to constrain the parameters of this model. Throughout this study, we adopt a pulsar braking index of $n=3$ (corresponding to magnetic dipole braking), a spindown timescale of $\tau=12 \, {\rm kyr}$, and normalize the total emission to match the flux observed by HAWC and HESS at multi-TeV energies. For the diffusion coefficient within 30 pc of Geminga, we adopt the following broken power-law parameterization:
\begin{equation}
    D(E_e) = \Biggl\{ \begin{array}{lll} 
    D_0 \left(E_e/{1 \,{\rm TeV}} \right)^{\delta_1}\quad &, \;\; E_e > 1 \, {\rm TeV} \\[3pt]
    D_0 \left(E_e/{1 \, \rm \text{TeV}} \right)^{\delta_2}\quad &, \;\; E_e < 1 \, \text{TeV},
    \end{array}
\end{equation}
where $D_0=10^{27} \, {\rm cm}^2/{\rm s}$ and $\delta_1 = 1/3$ are chosen to accommodate the spectrum observed at multi-TeV energies, as described in Ref.~\cite{Hooper:2023mvd} (see also, Refs.~\cite{Hooper_2017, Fang_2018, Profumo_2018, Tang_2019, J_hannesson_2019}). We treat $\delta_2$ as a free parameter which characterizes the evolution of the diffusion coefficient at sub-TeV energies. Beyond 30 pc from the pulsar, we adopt $D_0=4 \times 10^{29} \, {\rm cm}^2/{\rm s}$ and $\delta_1 = 1/3$, as inferred from measurements of the cosmic-ray boron-to-carbon ratio (and other secondary-to-primary ratios)~\cite{Strong:2007nh,Trotta:2010mx}. Note that 30 pc is comparable to the diffusion length of $\sim 10-100 \, {\rm GeV}$ electrons in the TeV halo, as integrated over the lifetime of Geminga. For this reason, we do not expect our results to change dramatically if the radius of the halo were larger than we have assumed.

In addition to variations in the value of $\delta_2$, we also consider different values for the spectral index of the injected electrons. Much like the diffusion coefficient, HAWC and HESS data constrain the value of this parameter at multi-TeV energies, but leave it largely unconstrained at lower energies, motivating a parameterization in the form of a broken power-law:
\begin{equation}
    \frac{dN_e}{dE_e} = \Biggl\{ \begin{array}{lll} 
    A \left(E_e/{1 \,{\rm TeV}} \right)^{-\alpha_1} \quad &, \;\; E_e > 1 \, {\rm TeV} \\[3pt]
    A \left(E_e/{1 \, \rm \text{TeV}} \right)^{-\alpha_2}\quad &, \;\; E_e < 1 \, \text{TeV},
    \end{array}
\end{equation}
where we adopt $\alpha_1=1.8$, set $A$ to match the intensity observed by HAWC~\cite{Hooper:2023mvd}, and take $\alpha_2$ to be a free parameter.

\begin{figure}[t]
    \centering
    \includegraphics[width=0.48\linewidth]{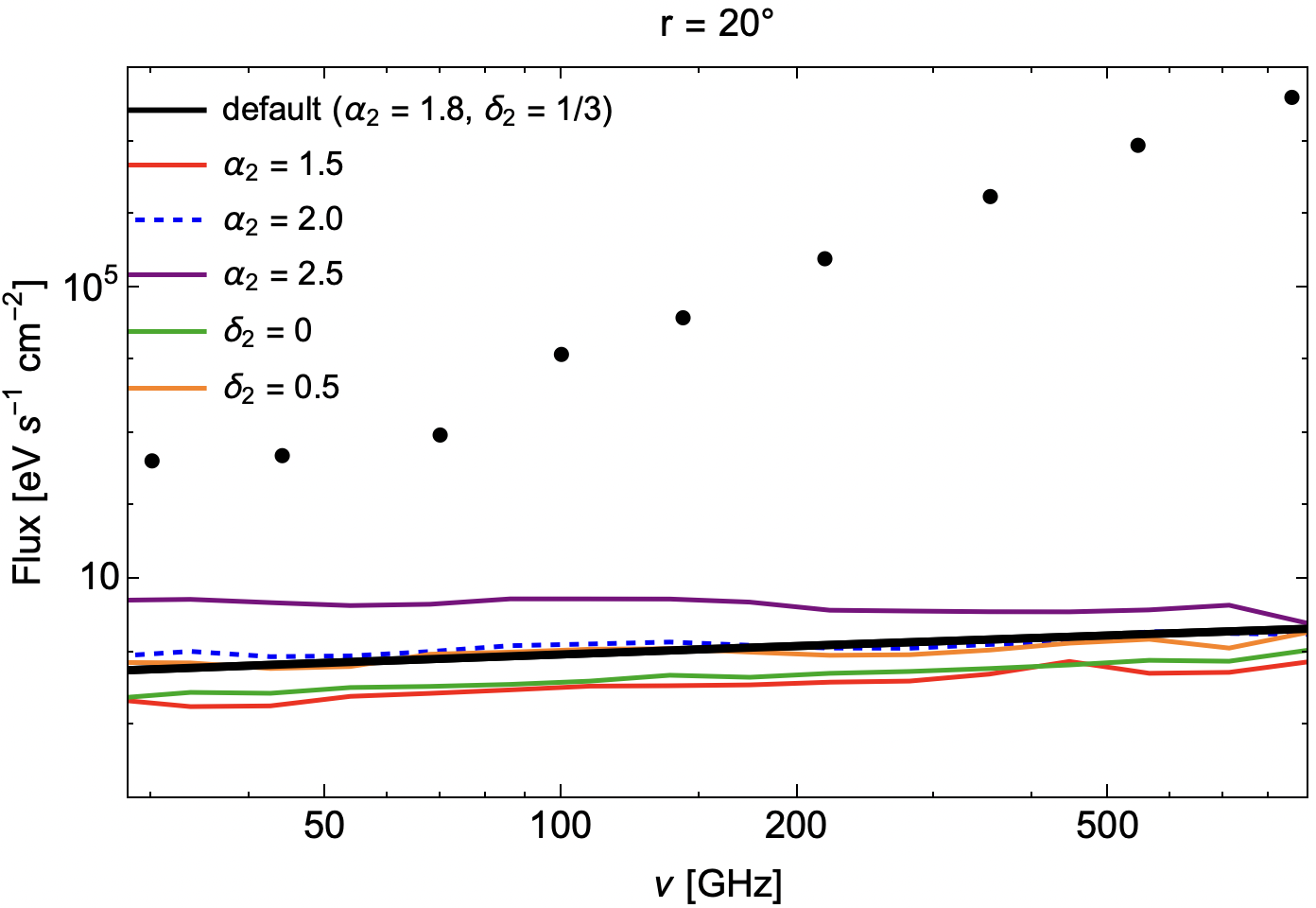}~\includegraphics[width=0.48\linewidth]{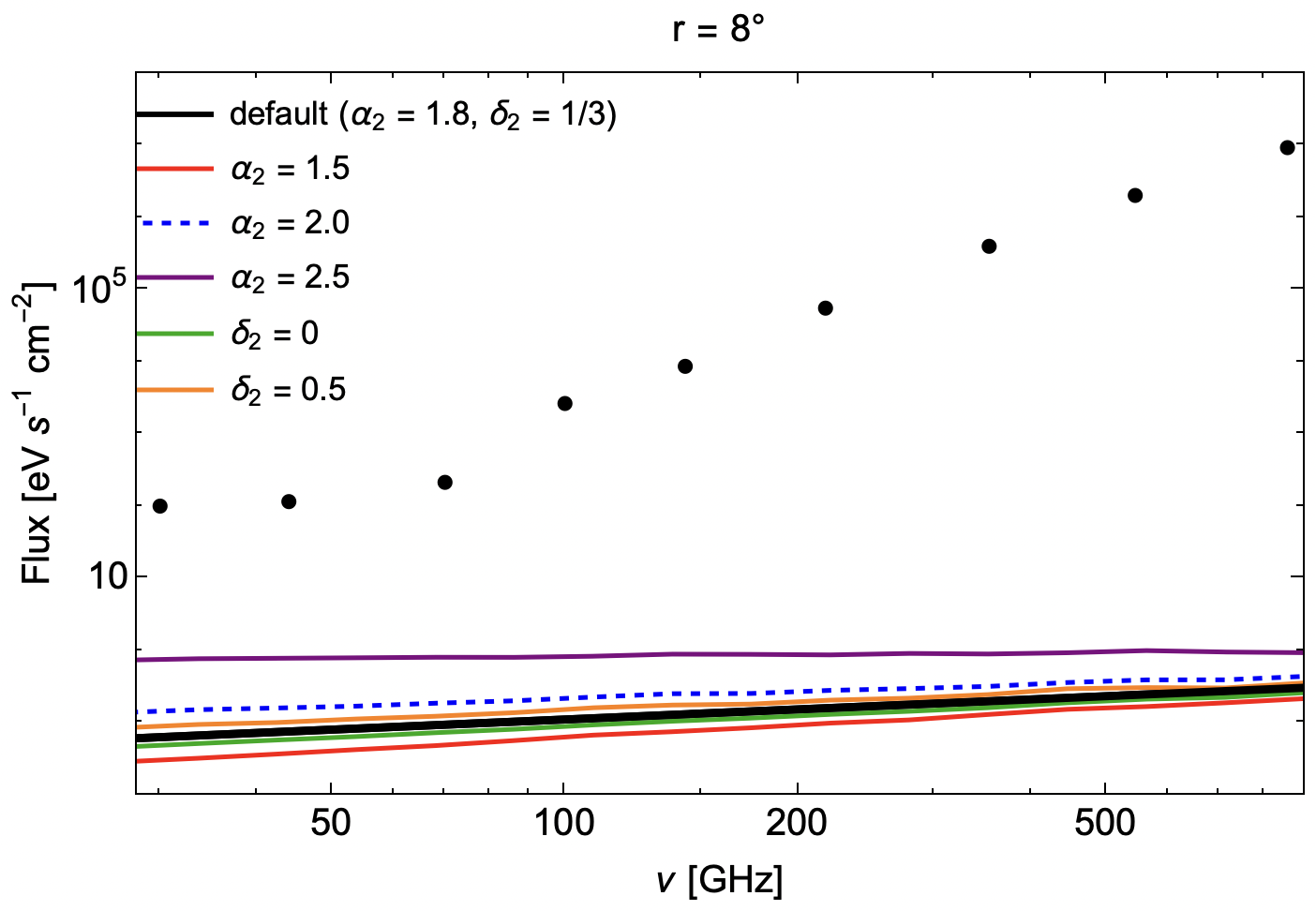}
    \caption{A comparison of the synchrotron spectrum predicted by our model of Geminga's TeV halo (lines) to the measurements of Planck (black points). For each of the model parameters we considered in this study, the predicted synchrotron flux is well below the total flux measured by Planck. The fluxes shown are integrated over a $20^{\circ}$ radius circle centered on the point ${\rm RA} = 91.6^\circ$, ${\rm DEC} = 12.8^\circ$ (left frame) and an $8^{\circ}$ radius circle centered on ${\rm RA} = 85.1^\circ$, ${\rm DEC} = 8.2^\circ$ (right frame).}
    \label{fig:flux} 
\end{figure}

Our main results are shown in Fig.~\ref{fig:flux}, where we compare the synchrotron spectrum predicted by our model to the measurements of Planck. These results are shown as integrated over two different regions: a $20^{\circ}$ radius circle centered on the point ${\rm RA} = 91.6^\circ$, ${\rm DEC} = 12.8^\circ$ (left frame) and an $8^{\circ}$ radius circle centered on ${\rm RA} = 85.1^\circ$, ${\rm DEC} = 8.2^\circ$, corresponding to Geminga birthplace where we expect the brightest synchrotron flux (right frame). For each of the model parameters we have considered (corresponding to different values of $\alpha_2$ and $\delta_2$), the predicted synchrotron flux is well below the total flux measured by Planck in each of their 9 frequency bands. These measurements are thus not capable of significantly constraining the values of these parameters at this time.

Note that we have not attempted to model or subtract any backgrounds from the Planck data. Such an analysis could potentially provide a stronger constraint on the model of Geminga's TeV halo.

\section{Summary and Conclusions}

Measurements by HAWC, LHAASO, and HESS  have shown that young and middle-aged pulsars convert a significant fraction of their total spin-down power into very high-energy electrons, leading to the formation of TeV halos. Although observations of pulsars at very high-energies can reveal a great deal about the nature of these objects, such measurements do not reveal much about the electrons that these sources might produce at lower energies, or how those particles propagate through the surrounding volume of space.

If pulsars efficiently accelerate electrons in $\sim 50-300 \, {\rm GeV}$ energy range, these particles would produce a spectrum of spatially extended synchrotron emission that peaks in the frequency range measured by Planck. Such emission could be used to constrain the low-energy diffusion coefficient in the regions surrounding these pulsars, as well as the spectrum and intensity of the electrons that are accelerated in this energy range.

In this study, we have attempted to use Planck data to constrain the nature of Geminga's TeV halo.  We find no conclusive evidence of emission in Planck's frequency range from this source, however, and calculate that the synchrotron flux from Geminga should be well below the total flux measured by Planck, even for models with favorable diffusion parameters or soft injection spectra. These measurements are thus not capable of significantly constraining the values of these parameters at this time.

In a spirit similar to that of the work presented here, the authors of Ref.~\cite{Manconi:2024wlq} have used X-ray measurements of Geminga to place an upper limit on the synchrotron emission from ultra-high-energy electrons around this source, reaching the conclusion that the magnetic field within Geminga's halo must be weaker than $\sim 2 \mu {\rm G}$. Data from the Fermi gamma-ray space telescope has also been used to search for the inverse Compton emission from GeV-TeV scale electrons within Geminga's halo~\cite{DiMauro:2019yvh}.

\section*{Acknowledgements}

We would like to thank Tom Crawford, Damiano Caprioli, and Benedikt Schroer for helpful discussions. The authors acknowledge support from the Fermi Research Alliance, LLC under Contract No.~DE-AC02-07CH11359 with the U.S.~Department of Energy, Office of High Energy Physics.

\appendix


\bibliographystyle{JHEP}
\bibliography{refs.bib}

\end{document}